# Exploring the OODA Loop as a Systematic Way of Thinking in Coping with Conflicts

Ethan Anderson[12] and Shouhuai Xu[1]

[1] Laboratory for Cybersecurity Dynamics Department of Computer Science University of Colorado Colorado Springs Colorado Springs, Colorado, USA 80918
[2] United State Air Force Academy Colorado Springs, Colorado, USA 80918

**Abstract.** When conflicts emerge, we need systematic ways of thinking to deal with them. This paper revisits Boyd's Observe–Orient–Decide–Act (OODA) loop and explores its usefulness as a systematic way of thinking for reasoning about conflicts in dynamic environments characterized by uncertainty, adaptation, and adversarial interference. We explore the OODA loop beyond its origin in air warfare where it focuses attention on the relationship between information, understanding, choice, and action. Our exploration is conducted in two application domains: cyber conflicts and cognitive conflicts. Across both domains, we emphasize *situational awareness* as a critical mechanism of orientation, while noting that observations become useful only when they are perceived, comprehended, projected into possible futures, and integrated with mental models, objectives, doctrine, trust, and experience. Our exploration suggests that conflict is not merely a contest of actions or effects, but a contest over the ability to generate, protect, and leverage one's own superior observation, orientation, and decision, while degrading and exploiting adversary's observation, orientation, and decision.

## 1 Introduction

Conflict forces an actor to reason under uncertainty while an adversary or opponent attempts to change both the environment and the actor's understanding of the environment. In such settings, a central question is not only what action should be taken at a given point in time, but also how an actor should think before acting (e.g., which action should one take?). A purely action-centric view of conflicts risks treating operations as isolated effects delivered against targets. This is far from sufficient because competent actors should have a decision-centric view, while noting that decision does not happen in isolation but is one component to the chain of actors' observation, interpretation, decision, and adaptation over time.

In the context of air warfare (or dogfight), Boyd introduced a compact way to express this decision-centric perspective [1], known as the Observe, Orient,

Decide, and Act (OODA) loop. At a high level, the loop can be understood as follows. An actor must collect signals from the environment, orient those signals within a changing mental and operational context, decide among possible courses of action, act, and then learn from feedback. The loop is often interpreted as a cycle of rapid action. However, its true value is much deeper and wider than this because of its account of adaptation, which can be extended to other kinds of conflicts than air warfare. Moreover, it paves a foundation for the systematic way of thinking in adversarial conflicts, where each actor is simultaneously trying to improve its own loop while degrading the adversary's loop.

This paper demonstrates and explores the wide applicability of the OODA loop through two domains in which conflict is increasingly important: the cyber domain and the cognitive domain, while noting that the former has been more intensively investigated than the latter. These two domains are related but distinct. On one hand, cyber conflicts concern operations through and against digital systems, networks, data, access, and computational infrastructure. On the other hand, cognitive conflicts concern operations that affect perception, comprehension, projection, trust, belief, narrative, and decision-making. Recent studies argue that cognitive warfare should be understood as a sustained and adaptive contest over human decision-making, rather than as a synonym for messaging, influence, or information dissemination [9, 11]. In both cyber and cognitive domains, however, the decisive issue is often not whether more data can be collected, but whether the actor can transform data into usable understanding before the adversary can corrupt, delay, overload, or exploit that understanding. The present study identifies what must be understood, compared, protected, or changed across competing OODA loops, rather than to prescribe how a particular operation should be executed. Accordingly, the discussion emphasizes mission conditions, relative decision advantage, operational choices, and observable out-comes. Technical methods matter, but they are treated as means that support the larger contest over awareness, orientation, and adaptation.

Although decision-making is core to the OODA loop, decisions do not happen in isolation. It is built on orientation, and orientation depends in part on situational awareness (SA). This makes observation important without reducing the paper to a sensor-centric discussion: observation provides the raw signals, SA makes those signals meaningful in relation to the current situation, and orientation places that awareness within the actor's objectives, assumptions, doctrine, experience, authorities, and constraints. Endsley's model defines SA in terms of *perception*, *comprehension*, and *projection* [6]. This model helps separate observation from understanding as follows: observation supplies data; SA makes relevant elements perceptible, meaningful, and projectable; orientation integrates that awareness with prior experience, doctrine, cultural assumptions, objectives, trust relations, and mental models. This suggests the following interpretation:

Observe → Orient accommodating Situational Awareness → Decide → Act,

which provides an explicit, rather than implicit, treatment of orientation.

**Our Contributions**. This paper makes three contributions. First, we advocate generalizing the OODA loop as a systematic framework for coping with conflicts.

This goes much beyond the original setting where the OODA loop was introduced (i.e., air warfare). Moreover, we articulate the use of OODA loop from both the offense and defense perspectives. Furthermore, we advocate treating each component of the OODA loop as a sub-mission, and articulate two key attributes for measuring the competency of OODA loop, namely *agility*, which reflects how rapid one OODA loop is, and *trustworthiness*, which reflects the trustworthiness of a OODA loop in terms of its observe, orient, decide components. To the best of our knowledge, we are the first to advocate treating observe, orient, decide, and act as sub-missions, and articulate such attributes to describe the competency or effectiveness of OODA loops.

Second, we advocate the explicit incorporation of SA (situational awareness) into the orientation component of the OODA loop. Moreover, we specify what kinds of information is needed for effective SA and thus effective orientation, and what kinds of information are relevant to orientation but go beyond SA. Information for effective SA includes indicators of the actor's own condition, the adversary's activity, the environment, mission dependencies, timing, and observed effects. Information beyond SA includes doctrine, rules of engagement, command intent, risk tolerance, prior experience, adversary models, cultural assumptions, organizational trust, and available authorities. This guides us to specify what kinds of data should be collected to support competent or effective orientation. This guides us to specify what kinds of data should be collected to support competent orientation.

Third, we demonstrate this through two case studies, one in the cyber conflict domain and one in the cognitive conflict domain. We apply the framework to cyber and cognitive conflicts from both defensive and offensive perspectives, showing that cyber warfare can be understood as competition over cyber terrain, while cognitive warfare can be understood as competition over awareness, meaning, trust, and orientation.

**Related Work**. The OODA loop has been widely adapted to the cybersecurity domain to guide cyber operations, as demonstrated by the Cybersecurity Dynamics framework [12–14], cloud cybersecurity architectures [7,16], and proactive cyber defenses [2]. Recently, the OODA loop has been adapted to the cognitive warfare domain [9, 11]. The present study goes beyond them in systematically exploring with emphasis on the orientation component of the OODA loop, while making explicit connection to the notion of SA, and to the observe component because the quality, timeliness, and relevance of observations bound what SA and orientation can achieve.

**Paper Organization**. Section 2 reviews the origin of the OODA loop, discusses its applications from both defense and offense perspective, and characterizes from two key aspects in agility and trustworthiness. Section 3 presents our case study for the cyber domain. Section 4 presents our case study for the cognitive domain. Section 5 concludes the paper with future research directions.

## 2 OODA Loop: Origin, Applications, and Characteristics

### 2.1 The Origin

Boyd's OODA loop [1] grew from the problem of prevailing in air-to-air combat (or dogfight), where two pilots continuously observe, interpret, decide, maneuver, and reassess under severe time pressure. The loop is not a simple four-step checklist. Instead, the loop depicts an ongoing interaction in which *observation* is shaped by unfolding circumstances and feedback, *orientation* is influenced by experience, analysis, synthesis, culture, and prior knowledge, *decision* selects a course of action from the actor's oriented understanding, objectives, constraints, and perceived opportunities, and *action* immediately changes what both participants can observe next. The loop is therefore relational: its value is revealed by comparing one's gain in terms of observation, orientation, decision, and action over the opponent's.

Putting into a more general context, the four components can thus be interpreted as a connected operational model. Observation collects signals about the environment, the opponent, friendly forces, and the effects of prior actions. Orientation transforms those signals into a mission-relevant understanding of the situation. Decision selects what condition or effect should be pursued. Action changes the environment and produces new feedback. In adversarial settings, every component is contested: observations may be concealed or falsified, orientation may be distorted, decisions may be delayed or forced, and actions may be disrupted or rendered irrelevant.

### 2.2 Applications

The OODA loop has two equally important perspectives: offense and defense, which are not necessarily explicitly presented in Boyd's original work as a formal taxonomy of separate offensive and defensive loops, but are implicit in the adversarial comparison at the center of [1]. These perspectives are particularly important when adapting the OODA loop to other conflicts.

**The Offense Perspective** From the offensive perspective of air warfare (or "dogfight"), the pilot seeks to create a fleeting position of advantage and convert it into a decision advantage (i.e., creating an orientation or decision problem the opponent cannot solve in time). The offensive pilot observes the opponent's energy, geometry, likely intent, and reaction; orients to where an advantage is emerging; decides whether the opportunity is real; and acts to increase the mis-match. The point is not movement for its own sake. It is to present changes, threats, and ambiguities that make the opponent's orientation increasingly obsolete. Offensive advantage exists when the opponent must react to a situation that has already changed.

**The Defense Perspective** From the defensive perspective, the pilot seeks to survive the immediate threat while preventing the opponent from preserving that advantage. The defender must recognize the developing problem, maintain awareness despite uncertainty and pressure, deny the attacker a predictable response, and create a transition back toward a neutral or favorable position. Defense is therefore about active adaptation, not passive absorption. A defensive maneuver matters because of what it changes in the relationship: the attacker's observation becomes less reliable, the attacker's orientation and decision become harder, and the defender gains another opportunity to observe, reorient, decide, and act to its advantage.

### 2.3 Characteristics

We define two key attributes to characterize the competency of both attack and defense OODA Loops. Each attribute is equally applicable to the four components of OODA.

**Agility**. Intuitively, agility describes how quickly a defender or attacker conducts a task (e.g., adapting to the opponent's adaptation) [3, 8, 15].

- **Observation Agility**. This attribute describes and measures how quickly the employed sensors can capture the data that reflect the opponent's new activities or maneuvers. If observation is treated as a sub-mission, this attribute measures how fast this sub-mission can be accomplished.
- **Orientation Agility**. This attribute describes and measures how quickly the orientation process can infer the desired information, such as the updated situational awareness. If orientation is treated as a sub-mission, this attribute measures how fast this sub-mission can be accomplished.
- **Decision Agility**. This attribute describes and measures how quickly decisions can be made. If decision is treated as a sub-mission, this attribute measures how fast this sub-mission can be accomplished.
- **Action Agility**. This attribute describes and measures how quickly the decision can be enforced or executed. If action is treated as a sub-mission, this attribute measures how fast this sub-mission can be accomplished.

**Trustworthiness**. Intuitively, trustworthiness describes the degree to which the outcome of a task can be trusted [3, 15].

- **Observation Trustworthiness**. This attribute describes and measures the degree the observations, or observed data, can be trusted for orientation and decision purposes. If observation is treated as a sub-mission, this attribute measures the degree this sub-mission is accomplished.
- **Orientation Trustworthiness**. This attribute describes and measures the degree the orientation outcome, such as situational awareness, can be trusted for decision purposes. If orientation is treated as a sub-mission, this attribute measures the degree this sub-mission is accomplished.
- **Decision Trustworthiness**. This attribute describes and measures the degree the decisions can be trusted for execution, while noting that execution incurs costs. If decision itself is treated as a sub-mission, this attribute measures the degree this sub-mission is accomplished.

- **Action Trustworthiness**. This attribute describes and measures the degree the action can achieve the intended objectives pertaining to the decision. If action is treated as a sub-mission, this attribute measures the degree this submission is accomplished.

## 3 OODA Loop in Cyber Domain

The cyber domain preserves the attacker–defender relationship even though its terrain, timing, and observables differ from a visual dogfight. A cyber attacker likewise seeks a position of advantage, attempts to keep the defender oriented to an incomplete or outdated picture, and adapts when the defender changes the environment. A cyber defender seeks to detect the developing mismatch, protect mission-relevant awareness, deny the attacker a stable understanding of the defended environment, and restore a favorable decision position. Unlike an air engagement, multiple cyber interactions may unfold simultaneously, remain partially hidden, and persist across tactical and operational timescales. Nevertheless, Boyd's central comparison holds: advantage belongs to the actor whose orientation remains more relevant to the changing conflict and who can use that orientation to shape what the other actor must confront next [1]. In what follows we adapt the OODA Loop way of thinking to defensive and offensive operations in the cyber domain, respectively.

### 3.1 The Defense OODA Loop in Cyber Domain

**Defensive Observation in Cyber Domain**. In defensive cyber operations, the defender employs various sensors to observe networks, endpoints, identities, applications, data flows, vulnerabilities, configurations, alerts, logs, external intelligence, and mission dependencies. The goal is not to collect everything, but to collect signals that can reveal adversary presence, intent, capability, opportunity, and likely next moves. Cyber defense often suffers from abundant data but limited understanding; therefore, observation must be organized around the mission and the adversary, not merely around available sensors.

High observation agility exists when a defender quickly notices that several weak signals across logs, endpoint telemetry, and network flows are appearing around a mission-critical planning system during an operationally significant window. Low observation agility exists when those same signals remain isolated in separate tools or queues until the adversary has already changed position. High observation trustworthiness exists when the collected observations are timely, correlated, and resistant to obvious spoofing or sensor blind spots. Low observation trustworthiness exists when the defender sees a large volume of alerts but cannot tell whether the alerts reflect adversary activity, benign administrative behavior, automated noise, or deliberate deception.

**Defensive Orientation in Cyber Domain**. The defender orients by trans-forming cyber observations into situational awareness and then integrating that awareness with mission priorities, threat models, defensive doctrine, asset criticality, known adversary tactics, and operational constraints. Cyber situational awareness includes perceiving relevant events, comprehending their significance, and projecting how an intrusion or campaign may evolve [4, 6]. A defender who sees an alert but cannot relate it to adversary objectives, key terrain, or mission risk has observed without orienting. Situational awareness is therefore a necessary but incomplete output of orientation. In cyber defense, orientation should also produce an assessment of mission impact, adversary intent, key cyber terrain, uncertainty, risk tolerance, response options, and expected adversary adaptation. Situational awareness answers what is happening and what may happen next; broader orientation answers what the situation means for the mission and what kinds of decisions are now sensible.

High orientation agility exists when the defender rapidly connects weak observations to a coherent hypothesis, such as recognizing that abnormal authentication attempts, unusual data staging, and reconnaissance of backup infrastructure are part of a campaign against mission continuity. Low orientation agility exists when the defender treats those observations as separate incidents and does not update the mission picture until after the adversary has shaped the environment. High orientation trustworthiness exists when the defender's assessment is grounded in validated telemetry, known mission dependencies, adversary trade-craft, and explicit uncertainty. Low orientation trustworthiness exists when the defender overfits the first plausible explanation, ignores contradictory evidence, or mistakes a decoy pattern for the adversary's main effort.

**Defensive Decision in Cyber Domain**. Defensive decisions include whether to monitor, contain, isolate, deceive, patch, reconfigure, restore, attribute, escalate, or accept risk. Decision quality depends on the trustworthiness of the defender's orientation. For example, containment may be appropriate when mission risk is immediate, while continued monitoring may be appropriate when preserving visibility into an adversary campaign is more valuable.

High decision agility exists when the defender can quickly choose an appropriate mission-protective response, such as isolating a compromised communication path while preserving essential command-and-control functions. Low decision agility exists when an organization understands the problem but waits for unclear authorities, excessive coordination, or unavailable approval chains. High decision trustworthiness exists when the chosen response follows from the orientation and contributes to accomplishing a mission, even if it accepts some operational risk (below a threshold). Low decision trustworthiness exists when the decision is fast but mismatched, such as shutting down a service that is more critical to the mission than the suspected adversary activity it was meant to stop.

**Defensive Action in Cyber Domain**. Defensive actions implement the selected course of action: blocking traffic, disabling accounts, deploying detection logic, segmenting networks, moving services, restoring backups, changing credentials, engaging deception assets, or communicating with commanders and stakeholders. These actions should also be treated as probes that generate feed-back. A good defensive action not only changes the environment but improves the defender's next observation and orientation.

High action agility exists when the defender can execute the decision quickly through prepared playbooks, validated access, trained personnel, and resilient technical controls. Low action agility exists when a good decision cannot be implemented because the team lacks privileges, tooling, current asset knowledge, or rehearsed procedures. High action trustworthiness exists when the action has the intended operational effect and produces useful feedback for the next observation and orientation cycle. Low action trustworthiness exists when the action appears successful in a tool but does not actually change the adversary's position, creates unanticipated mission harm, or blinds the defender to the next phase of the campaign.

## 3.2 The Offense OODA Loop

**Offensive Observation in Cyber Domain**. In offensive cyber operations, the attacker observes exposed services, vulnerabilities, identities, trust relationships, network topology, defensive tools, operational rhythms, mission dependencies, and user behavior. Reconnaissance is not merely a pre-attack phase; it continues throughout an operation as the attacker learns how the defender changes.

For an offensive actor, high observation agility exists when reconnaissance quickly detects changes in the defense posture, such as newly deployed monitoring, altered access controls, changed user behavior, or a patched exposure. Low observation agility exists when the attacker continues to rely on a stale picture of the environment after the defender has maneuvered. High observation trustworthiness exists when the attacker can distinguish real defender changes from noise, honeypots, and deception. Low observation trustworthiness exists when the attacker mistakes a planted signal or incomplete scan result for the true terrain and orients on a false opportunity.

**Offensive Orientation in Cyber Domain**. The attacker orients by interpreting observations as opportunities and constraints. This includes identifying key cyber terrain, estimating defender visibility, assessing access durability, mapping likely response paths, and choosing how to preserve or improve position. A maneuver-centric view of cyberspace emphasizes gaining, maintaining, and exploiting positions of advantage rather than treating cyber merely as a fires capability that delivers discrete effects [5]. In OODA terms, cyber maneuver depends on effective orientation: the attacker seeks a better understanding of the terrain and a more advantageous position within it.

High offensive orientation agility exists when the attacker rapidly interprets observed defender changes and updates the operation, for example by recognizing that a visible defensive response is protecting one mission dependency while leaving another dependency exposed. Low offensive orientation agility exists when the attacker keeps pursuing an access path whose value has already been reduced. High offensive orientation trustworthiness exists when the attacker's assessment of terrain, access, defender visibility, and mission relevance is accurate enough to support the next decision. Low offensive orientation trustworthiness exists when the attacker misjudges the defender's visibility or mistakes a low-value system for key terrain.

**Offensive Decision in Cyber Domain**. Offensive cyber decisions include which access vector to use, whether to exploit or preserve a vulnerability, whether to move laterally or remain dormant, whether to continue or abort an operation, and whether to pursue collection, disruption, deception, or preparation of the environment. These decisions depend on the attacker's assessment of opportunity, risk, timing, and expected defender response.

High offensive decision agility exists when the attacker can quickly choose whether to preserve access, shift objectives, exploit a fleeting opportunity, or terminate an operation after the defender changes conditions. Low offensive decision agility exists when the attacker has observations and orientation but cannot decide before the window closes. High offensive decision trustworthiness exists when the selected course of action improves the attacker's position relative to the mission objective and expected defender response. Low offensive decision trustworthiness exists when the attacker chooses a tempting action that reveals position, burns access, or produces an effect that does not contribute to the higher objective.

**Offensive Action in Cyber Domain**. Offensive actions include scanning, phishing, exploiting, establishing persistence, escalating privileges, moving laterally, exfiltrating data, disrupting services, manipulating data, or creating false signals. These actions also attempt to shape the defender's OODA loop by hiding true activity and intent, producing misleading observations, overloading analysts, or forcing decisions under time pressure. In this sense, cyber fires and cyber maneuver differ in emphasis [5]: fires seek an effect against a target, whereas maneuver seeks a continuing position of advantage that can improve one's own OODA loop and pressure the adversaries.

High offensive action agility exists when the attacker can execute the selected course of action before the defender's new orientation closes the opportunity. Low offensive action agility exists when the attacker decides correctly but cannot act because access is fragile, tooling is unsuitable, or timing has passed. High offensive action trustworthiness exists when execution produces the intended effect while preserving the attacker's ability to observe and adapt. Low offensive action trustworthiness exists when execution fails, creates unintended indicators, alerts the defender prematurely, or changes the environment in a way that damages the attacker's own follow-on options.

## 4 OODA Loop in the Cognitive Domain

The cognitive domain concerns conflict over how people and organizations perceive, comprehend, project, trust, and decide. In this domain, the object of contest is not only a network, a system, or an access path, but the process that allows an actor to understand what is happening and what response is appropriate.

Cognitive conflict is therefore well suited to an OODA-based treatment because it directly targets the relationship between observation, orientation, decision, and action. Recent work argues that cognitive warfare should be understood as a sustained contest over decision-making rather than messaging, persuasion, or information dissemination [9, 11]. The relevant question is what must be protected, recognized, compared, or changed in the cognitive environment so that friendly actors preserve agile and trustworthy orientation while adversaries lose decision advantage.

### 4.1 Defense OODA Loop in Cognitive Domain

**Defensive Observation in Cognitive Domain**. In defensive cognitive operations, the defender observes information flows, narratives, audience reactions, trust relationships, media artifacts, coordinated behavior, source provenance, sentiment shifts, and indicators of manipulation. Observation must include both content and context: who is communicating, through which channels, to which audiences, at what time, with what apparent reach, and with what apparent decision effect. Because cognitive conflict often develops through accumulation rather than a single event, the defender should also observe changes in confidence, confusion, trust calibration, decision latency, coordination, and willing-ness to act.

High observation agility exists when the defender quickly detects that a narrative, media artifact, or coordinated behavior is emerging around a mission-relevant audience before it shapes downstream decisions. Low observation agility exists when the defender notices the activity only after the audience has already incorporated it into its understanding of the situation. High observation trust-worthiness exists when the defender can distinguish authentic audience concern from coordinated manipulation, platform noise, satire, organic disagreement, or adversary deception. Low observation trustworthiness exists when the defender sees volume or virality but cannot assess source, coordination, audience penetration, or mission relevance.

**Defensive Orientation in Cognitive Domain**. The defender orients by interpreting how observed information activity affects perception, comprehension, projection, trust, and decision-making. Cognitive situational awareness includes understanding what relevant audiences are seeing, what they appear to believe, how they may project the situation forward, and how those projections may affect decisions. Broader orientation goes beyond situational awareness by incorporating audience identity, cultural context, prior beliefs, institutional trust, adversary objectives, command intent, legal and policy constraints, likely second order effects, and the risk that a defensive response may amplify the harmful activity.

High orientation agility exists when the defender rapidly connects observed information activity to a plausible adversary purpose and mission consequence. For example, a defender may recognize that a false story about disrupted logistics is not merely a rumor but an attempt to undermine confidence in an upcoming operation. Low orientation agility exists when the defender treats the same activity as isolated content moderation, public affairs, or social media noise rather than as part of an adversarial campaign against decision-making. High orientation trustworthiness exists when the defender's assessment is grounded in source provenance, audience analysis, behavioral indicators, mission context, and explicit uncertainty. Low orientation trustworthiness exists when the defender assumes intent without evidence, confuses popularity with influence, or fails to distinguish a narrative's existence from its effect on decisions.

**Defensive Decision in Cognitive Domain**. Defensive decisions include whether to monitor, ignore, debunk, attribute, amplify trusted sources, engage community leaders, change communication channels, coordinate with platforms, or take technical action against coordinated manipulation. The defender must decide not only whether the information is false or harmful, but what condition should be created in the audience's OODA loop: reduced confusion, restored trust, better source discrimination, preserved confidence, or delayed adversary exploitation.

High decision agility exists when the defender can quickly choose a response that matches the cognitive condition at risk, such as using a trusted messenger before a misleading narrative becomes the dominant frame. Low decision agility exists when the defender understands the problem but delays because authorities, ownership, messaging, or inter-agency coordination are unclear. High decision trustworthiness exists when the chosen response protects the audience's ability to maintain sound orientation without creating greater harm. Low decision trustworthiness exists when the response is fast but counterproductive, such as amplifying a marginal narrative, appearing defensive in a way that re-duces trust, or addressing factual accuracy while ignoring the audience's deeper concern.

**Defensive Action in Cognitive Domain**. Defensive actions include public communication, trusted messenger participation, media literacy support, platform coordination, transparency measures, community participation, and institutional response. The goal is not merely to remove content, but to protect the audience's ability to observe accurately, orient soundly, decide appropriately, and act in accordance with mission-relevant reality. Cognitive war gaming can help rehearse such actions under adversarial pressure, especially when exercises represent realistic audiences, attacker-defender interaction, decision effects, and adaptive adversary behavior [10].

There is high action agility when the defender can execute a selected response through prepared communication channels, trusted relationships, approved authorities, and rehearsed coordination. Low action agility exists when a good decision cannot be executed before the adversary narrative hardens or spreads to new audiences. High action trustworthiness exists when the action produces the intended cognitive condition, such as improved source discrimination, reduced uncertainty, preserved confidence, or restored alignment between audience understanding and operational reality. Low action trustworthiness exists when the action reaches the wrong audience, lacks credibility, triggers backlash, or prevents the defender from observing how the audience actually responded.

## 4.2 The Offense OODA Loop

**Offensive Observation in Cognitive Domain**. An offensive actor observes audiences, grievances, identities, trust networks, cultural narratives, media habits, institutional weaknesses, crisis conditions, and existing uncertainty. The attacker seeks to understand where an audience's perception, comprehension, projection, or trust can be manipulated to create decision advantage. As in cyber operations, observation is not only a preparatory phase; the attacker continues observing how the audience and defender respond.

High offensive observation agility exists when the attacker quickly detects a change in audience mood, institutional credibility, or operational context that creates a new opening for influence. Low observation agility exists when the attacker continues pushing a narrative after the audience has moved on or the defender has changed the information environment. High offensive observation trustworthiness exists when the attacker can distinguish real audience vulnerability from platform artifacts, bot activity, misleading polling, or wishful assumptions. Low observation trustworthiness exists when the attacker misreads the audience and orients on a grievance, symbol, or channel that is not actually decision-relevant.

**Offensive Orientation in Cognitive Domain**. The attacker orients by identifying cognitive terrain: narratives, communities, influencers, symbols, emotional triggers, trust relationships, uncertainty, and decision points. This is analogous to cyber terrain in a maneuver-centric cyber framework [5]. Whereas cyber maneuver seeks positions of advantage in networks and systems, cognitive maneuver seeks positions of advantage in the adversary's sense-making process. The attacker may pursue acute effects that immediately disrupt observation, orientation, decision, or action, or chronic effects that slowly alter trust, priors, and interpretive baselines [9].

High offensive orientation agility exists when the attacker rapidly interprets audience feedback and adapts the campaign, for example by shifting from a failed factual claim to a more effective frame that exploits uncertainty or distrust. Low offensive orientation agility exists when the attacker continues using a theme that no longer resonates or that the defender has successfully interpreted. High offensive orientation trustworthiness exists when the attacker accurately understands which audience, trust relationship, or decision point is most vulnerable. Low offensive orientation trustworthiness exists when the attacker mistakes reach for effect, misjudges audience identity, or fails to anticipate that the campaign will trigger resistance rather than confusion.

**Offensive Decision in Cognitive Domain**. Offensive decisions include which audience to target, which narrative to promote, whether to deceive or selectively frame true information, whether to induce confusion, distrust, urgency, apathy, polarization, or overconfidence, and whether to pursue immediate disruption or longer-term conditioning. These decisions aim to degrade the defender's or audience's OODA loop by corrupting observation, distorting orientation, forcing poor decisions, or triggering counterproductive action.

High offensive decision agility exists when the attacker can quickly choose the cognitive effect that best exploits the current audience condition. Low decision agility exists when the attacker observes an opening but cannot select a timely narrative, channel, or audience before the condition changes. High offensive decision trustworthiness exists when the selected line of effort advances the higher objective by affecting decisions, not merely by generating attention. Low decision trustworthiness exists when the attacker chooses an attention-grabbing action that produces visibility but no meaningful decision advantage, or that causes the target to become more resilient.

**Offensive Action in Cognitive Domain**. Offensive actions include misinformation, disinformation, impersonation, deepfakes, narrative laundering, selective amplification, harassment, coordinated inauthentic behavior, information flooding, and manipulation of provenance or trust cues. Such actions can attack the OODA loop at multiple points: they can pollute observation, corrupt situational awareness, distort orientation, force poor decisions, or trigger counterproductive actions. The distinguishing issue is not the specific medium employed, but whether the activity produces relative decision advantage by degrading the target's OODA performance.

High offensive action agility exists when the attacker can execute the selected activity while the target audience is still susceptible and before the defender can reorient. Low action agility exists when execution occurs after the moment has passed or after the defender has inoculated the audience. High offensive action trustworthiness exists when the action produces the intended cognitive effect while preserving the attacker's ability to observe, adapt, and continue the campaign. Low action trustworthiness exists when the action is exposed, attributed, rejected by the audience, or so poorly aligned with the audience's context that it strengthens the defender's credibility.

## 5 Conclusion

This paper explores the OODA loop as a systematic way of thinking for coping with conflicts in domains where uncertainty, adaptation, and adversarial interference are central. Boyd's original formulation emerged from air warfare, but its deeper value is not confined to dog-fighting or to speed alone. Its value lies in comparing competing actors' abilities to observe relevant conditions, orient those observations into mission-relevant situation understanding, decide under uncertainty, act effectively, and learn from the feedback produced by action.

The paper makes three claims. First, the OODA loop should be treated as a decision-centric framework rather than merely as a cycle of rapid activity. Second, each component of the loop can be treated as a sub-mission whose competency can be characterized by agility and trustworthiness. Third, situational awareness is a critical mechanism of orientation, but orientation also includes information that goes beyond situational awareness, including doctrine, intent, authority, trust, experience, adversary models, and mission context. These dis-tinctions help an operator ask what must be observed, what must be understood, what decision condition must be created, and what action must be trusted to accomplish a higher mission.

The cyber and cognitive case studies show that the same framework can be applied across different forms of conflict without collapsing their differences. In cyber conflict, actors compete over visibility, access, terrain, timing, mission dependencies, and positions of advantage. In cognitive conflict, actors compete over perception, comprehension, projection, trust, meaning, and decision confidence. In both cases, the central issue is relative decision advantage: whether one actor can preserve a more agile and trustworthy OODA loop while degrading, delaying, misleading, or exploiting the adversary's loop.

This perspective is intentionally focused on the "what" rather than the "how." It does not prescribe specific techniques for cyber or cognitive operations. Instead, it identifies the operational conditions that matter: what observations are needed, what orientation must produce, what decisions must accomplish, and what actions must change in the environment. Future work should refine quantitative and qualitative measures of OODA agility and trustworthiness, develop richer cyber and cognitive case studies, and explore how operators can compare friendly and adversary OODA loops across tactical, operational, and strategic time horizons.